\documentclass[10pt,conference, final, letterpaper]{IEEEtran}
\usepackage{cite}
\usepackage{amsmath,amssymb,amsfonts}
\usepackage{graphicx}
\usepackage{textcomp}
\usepackage{gensymb}
\usepackage{tablefootnote}
\usepackage{algpseudocode}
\usepackage{algorithm}
\usepackage{xcolor}
\usepackage{caption}
\usepackage{subcaption}
\graphicspath{{./images/}}

\def\BibTeX{{\rm B\kern-.05em{\sc i\kern-.025em b}\kern-.08em
		T\kern-.1667em\lower.7ex\hbox{E}\kern-.125emX}}
\title{Distributed Probabilistic Congestion Control in LEO Satellite Networks}
\author{\IEEEauthorblockN{Pranav S. Page\IEEEauthorrefmark{1}, Kaustubh S. Bhargao\IEEEauthorrefmark{2}, Hrishikesh V. Baviskar\IEEEauthorrefmark{3}, Gaurav S. Kasbekar\IEEEauthorrefmark{4}} 
\IEEEauthorblockA{Department of Electrical Engineering\\ Indian Institute of Technology Bombay\\
Email : \IEEEauthorrefmark{1}pranavpage@ee.iitb.ac.in, \IEEEauthorrefmark{2}kbhargao@ee.iitb.ac.in,
\IEEEauthorrefmark{3}18d070048@iitb.ac.in,
\IEEEauthorrefmark{4}gskasbekar@ee.iitb.ac.in
}}
\renewcommand{\baselinestretch}{0.83}
\begin{document}
\maketitle	
	\begin{abstract}

In a dense Low Earth Orbit (LEO) satellite constellation, using a centralized algorithm for minimum-delay routing would incur significant signaling and computational overhead. In this work, we exploit the deterministic topology of the constellation to calculate the minimum-delay path between any two nodes in a satellite network. We propose a distributed probabilistic congestion control scheme to minimize end-to-end delay, which is built on top of the existing Datagram Routing Algorithm (DRA). The decision to route packets is taken based on the latest traffic information received from neighbours. We provide an analysis of the congestion caused by a simplified DRA on a uniform infinite mesh of nodes. We compare the proposed congestion control mechanism with the existing congestion control used by the DRA via simulations, and show improvements over the latter.

	\end{abstract}
\section{Introduction}

Dense Low Earth Orbit-based communication networks \cite{spacex, barnett2018oneweb, iridium} have become operational and are being preferred over geostationary satellites due to the lower ground-to-satellite propagation delay\cite{jamalipour}. Challenges faced by satellite constellations differ from those faced by terrestrial networks. Due to the dynamic and fast-moving nature of the constellation relative to the ground, association, handover and pointing between satellites are non-trivial problems\cite{handover, pointing}. The inter-satellite links are characterized by high propagation and transmission delays and high bit error rates \cite{isl_perf}. The satellites themselves have limited storage and processing capabilities \cite{trends_sats}, resulting in dropped packets in congested parts of the network. A centralized algorithm for routing would need a lot of transmissions and computation in order to send optimal paths to nodes in the network. The Datagram Routing Algorithm (DRA) exploits the geometry of the network and calculates the optimum minimum-delay path between a given pair of nodes. After that, it is the job of the congestion control algorithm to pick the next hop for a packet to reach its destination with the minimum queuing and propagation delay. The problem of choosing the next hop for the packet in the presence of congestion is the problem that this work focuses on. The choice has to be made locally, and without knowledge of the congestion level of every node along the minimum delay path.
A distributed congestion control algorithm that can deal with uneven node congestion levels to route packets from source to destination with low packet drops and end-to-end delay would be easy to implement on-board, and would offer better Quality of Service (QoS).

The problem of choosing an optimum schedule that minimizes the total end-to-end delay for a given set of packets has been shown to be NP-hard in \cite{opt_schedule}. Also, the problem of deciding whether a schedule exists whose completion time is bounded by some fixed given value $k \geq 0$ is proved to be NP-complete in \cite{bovet}. In \cite{hardness_opt_schedule}, it is shown that approximating the minimum end-to-end delay is NP-hard. Thus, heuristic-based approaches have been used to tackle this problem. Werner et al.\cite{atm-based} proposed a connection-oriented routing algorithm for meeting QoS requirements, which might not be able to handle link failures well. Kor\c{c}ak et al.\cite{epar} uses a priority metric to decide between multiple shortest paths in the network. In \cite{subopt}, the authors propose running a modified Dijkstra's algorithm on topology snapshots. Bertsekas \cite{bert}  discusses an asynchronous distributed algorithm for a broad class of dynamic programming problems that is guaranteed to arrive at the optimal estimate stored in each node as the computation proceeds infinitely. Link state broadcasting is used in \cite{breadth_first}, on which the breadth first tree search is performed. 
Distributed geographical routing is performed in \cite{traffic_classification}, and traffic is classified according to delay sensitivity.In \cite{survivable_inclined}, the focus is on distributed routing and it uses a restricted flooding approach to deal with link failures. The Low Complexity Probabilistic Routing algorithm in \cite{LCPR} describes a probabilistic routing technique for polar orbits with a simpler priority metric, which is a form of load balancing being done on the basis of congestion level information.

To deal with sudden topology changes, some form of link-state broadcast is needed. Marcano et al. \cite{AODV} propose an Ad hoc On-Demand Distance Vector routing protocol, and compare its performance with a flooding protocol to study its performance in terms of packet throughput. Luo et al. \cite{ref_dijkstra} introduce a refined Dijkstra's algorithm to deal with multiple shortest paths that are generated by the classic Dijkstra's algorithm run on the network.

Liu et al. \cite{load_balancing} propose Selective Shunt Load Balancing (SSLB), which is a distributed congestion control algorithm, using the traffic flow as an estimate of congestion in a part of the network. Dong et al. \cite{extended_link_states} introduce a load balancing algorithm based on extended link states to deal with congestion. 
Zhang et al. \cite{aser} propose an Area-based Satellite Routing algorithm, which is a distributed routing protocol dividing the LEO network into areas, and thus segregating the routing decisions into inter-area routing and intra-area routing. Huang et al. \cite{adaptive_multipath} use a pheromone based algorithm, called the Adaptive Multipath Traffic Scheduling algorithm (AMTS), which is a strategy using pheromones as indicators of network state, which is divided into node state and link state. Zhang et al. \cite{link_state} propose an on-board routing algorithm based on link state information received from neighboring nodes. 

Ekici et al. \cite{ekici-datagram, ekici-dist} propose the DRA which routes packets between virtual nodes based on their relative location. The DRA uses a basic threshold on the outgoing buffer to determine whether the link is congested, rather than using local congestion information. We address this problem by using the packet headers as a way of conveying traffic information in the form of a single metric indicating the congestion level, and then probabilistically choosing the next hop for a packet. We perform a mathematical analysis of the congestion control algorithm in a simple network. We simulate a typical LEO satellite constellation, and compare the performance of the DRA and our own algorithm in terms of end-to-end delay. In particular, we show that for a congested portion of the network, the probabilistic congestion control algorithm offers consistently low average end-to-end delay across flows, under high loads.

\section{System Model and Problem Formulation}
\label{sec:sys_model}
\begin{figure}
	\centering
	\includegraphics[width=0.4\columnwidth]{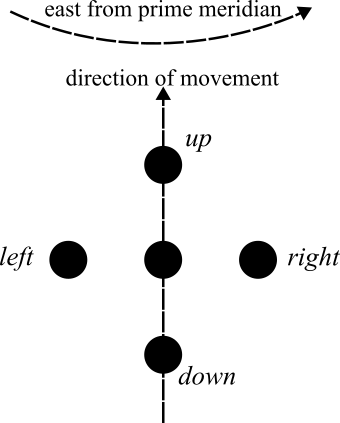}
	\caption{Neighbors of a node with given orientation with respect to prime meridian and given direction of movement}
	\label{fig:positions}
\end{figure}

The LEO satellite constellation considered is a Walker star \cite{walker} constellation, with satellites in polar orbits. The setup and terminology used is based on the terminology used in \cite{ekici-datagram}. There are $N$ orbital planes, with $M$ satellites per plane. Thus, the angular spacing between the planes is ${360^\degree}/{2N}$. \begin{figure}
	\centering
	\includegraphics[width=.7\columnwidth]{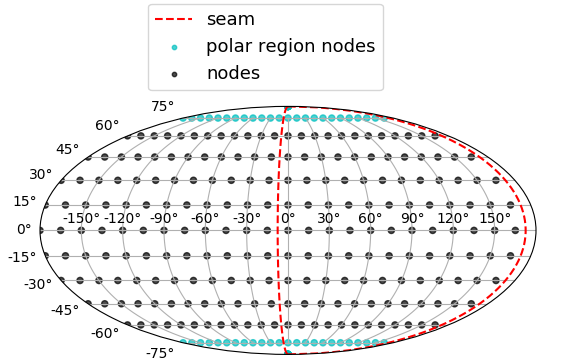}
	\caption{Satellite constellation with $N=12, M=24$ and counter-rotating seam indicated}
	\label{fig:constellation}
\end{figure}
 All satellites are at a fixed altitude $h$ from the ground, thus forming an orbital shell. We consider the network to be comprised of \textit{virtual nodes}, as in \cite{ekici-datagram}, with different satellites occupying the virtual nodes at different instances of time. The virtual nodes are fixed in location with respect to the ground, and are filled up by the nearest satellite. Each virtual node location can be expressed in terms of the plane number $p$ and the satellite number $s$, $0\leq p \leq N-1$, $0\leq s \leq M-1$. Each node in the network can be represented as a tuple $(p,s)$. This model does not deal with the mobility of satellites, and can be used to perform routing based entirely on the position of the virtual nodes, although satellites may not be at their exact locations due to orbital movement. Ekici et al. \cite{ekici-dist} provide results showing that this approximation has little effect on the end-to-end delay of packet flows, as the movement is small in magnitude. The directions of the neighbors are described in Figure \ref{fig:positions}.

 Inter-plane Inter Satellite Links (ISLs) in the polar regions are considered to be shut off due to the change in the orientation of neighbors. Polar regions are defined using a latitude threshold $\theta_{\text{polar}}$, with $\theta_{\text{polar}} = 75^\degree$ taken as the default boundary. The latitudes above $\theta_{\text{polar}}$ are considered to be in the polar regions.

Each node has four output buffers, corresponding to its four neighbors. We assume that on one link, reception and transmission can take place simultaneously. Thus, all four antennas can be transmitting and receiving simultaneously.

As seen in Figure \ref{fig:constellation}, the inter-plane ISLs are shorter towards the poles and longer towards the equator. The lengths of the intra-plane ISLs are the same, as nodes are equally spaced in the orbital planes. The lengths of the inter-plane ISLs decrease with increasing latitude. 

Now that the system model has been defined, the problem statement reduces to the following: Given the topology of the network of nodes and a destination node, without any global knowledge of the congestion level at each node, pick the next hop for the packet from the source node in a way that the end-to-end delay ($d_{\text{prop}}+d_{\text{queueing}}$) is minimized, where $d_{\text{prop}}$(respectively, $d_{\text{queueing}}$) is the total propagation (respectively, queueing) delay  encountered by the packet while going from source to destination.

\section{Algorithms}
\label{sec:algorithms}
The DRA consists of three phases, namely direction estimation, enhancement and congestion control. To compare the proposed congestion control algorithm with the DRA congestion control, an optimal path is obtained from the first two phases, and then the different congestion control algorithms are compared with respect to end-to-end delay. The parameters describing a path are $(n_h,n_v)$ (number of hops in horizontal/vertical directions) and $(d_h, d_v)$ (orientation in horizontal/vertical directions)
\subsection{Routing}
\label{sec:routing}
\subsubsection{DRA Direction Estimation}
In direction estimation, the path from source to destination with the minimum number of hops is chosen across paths crossing the polar region and not crossing the polar region.

\subsubsection{DRA Direction Enhancement}
This phase actually takes into account the unequal interplane ISLs in the network, and the transport of packets in the polar region. In this phase, the directions given by $d_h$ and $d_v$ are labeled primary or secondary, depending on the position of the packet source and destination. The primary and secondary directions are passed to the congestion control algorithm.

\subsection{Congestion Control}
\subsubsection{DRA Congestion Control}
In DRA, the decision to send packets in the primary or secondary directions is taken on the basis of the congestion level of the node's output buffers in the respective directions. If the output buffer in the primary direction has less than $N_{threshold}$ packets, then the packet is sent in the primary direction. Otherwise, if the secondary direction exists (might not exist for nodes in polar regions), and the output buffer in the secondary direction has less than $N_{threshold}$ packets, then the packet is sent in the secondary direction. If the buffers in both primary and secondary directions have buffers of size greater than $N_{threshold}$, then the packet is sent in the primary direction.
\subsubsection{New Congestion Control}

The proposed congestion control algorithm takes advantage of the dense network of nodes to transmit traffic information about a node to its neighbor. With each packet sent to its neighbor, the node adds a traffic metric to the header of the packet. Similarly, when a packet is received, the traffic metric in the header is extracted and stored in each node. Thus, each node maintains an output buffer and the latest traffic metric received for each of its neighbors. The decision to be taken involves the lengths of the output buffers as well as the traffic metric in each of the two directions.

The traffic metric to be sent has to be indicative of the congestion level of the node. The following weighted sum is used:
\begin{equation}
\label{eq:weighted_sum}
	m_{node} = \frac{\sum_{i=1, i\neq j}^4 (w_{ngbr}m_{node,i}+ (1-w_{ngbr})N_{node, i}) }{3}
\end{equation}
where $m_{node}$ is the traffic metric to be sent from $node$ to the neighbor $j$, $w_{ngbr}\in [0,1]$ is the weight given to the values of the traffic metric of the neighbors of the $node$, $N_{node,i}$ is the length of the output buffer in $node$ towards neighbor $i$, and $m_{node, i}$ is the latest traffic metric received by $node$ from neighbor $i$.  

Let $primary$ and $secondary$ be the primary and secondary directions given by the direction enhancement algorithm. If $N_{node, primary}$ and $N_{node,secondary}$ are the output buffer lengths in the respective directions, and $m_{node, primary}$ and $m_{node, secondary}$ are the traffic metrics received from the primary and secondary directions and stored in $node$, then the following congestion level metrics are calculated
\begin{equation}
\label{eq:w_buffer}
	c_{i} = w_{buffer}N_{node,i} + (1-w_{buffer})m_{node, i}
\end{equation}
for $i \in \{primary, secondary\}$, $w_{buffer}\in [0,1]$. A probability distribution is chosen such that 
\begin{align*}
	P(\text{choose primary}) &= f_{p}(c_p, c_s, p_{pref})\\ &:=\dfrac{(c_{s}+1)p_{pref}}{c_{p}+1 + (c_{s} - c_{p})p_{pref}}
\end{align*}
where $c_s = c_{secondary}$, $c_p = c_{primary}$ are the congestion level metrics in the respective directions, and $p_{pref}$ is the probability that primary is chosen when $c_s = c_p$. Similar to the DRA congestion algorithm, if $c_{p}<N_{threshold}$, then the packet is sent along the primary direction. If not, then the packet is sent in the primary direction with probability $P(\text{choose primary})$. 
The probabilistic nature of this choice ensures that a single direction is not clogged, which can happen in the DRA congestion control algorithm. Also, this ensures that the primary direction is not completely abandoned if it is clogged; the algorithm sends packets to it at a reduced rate. Preference is still given to the primary direction. 

\subsection{Mathematical Analysis}
\label{sec:math}

\begin{figure}
	\centering
	\includegraphics[width=.6\columnwidth]{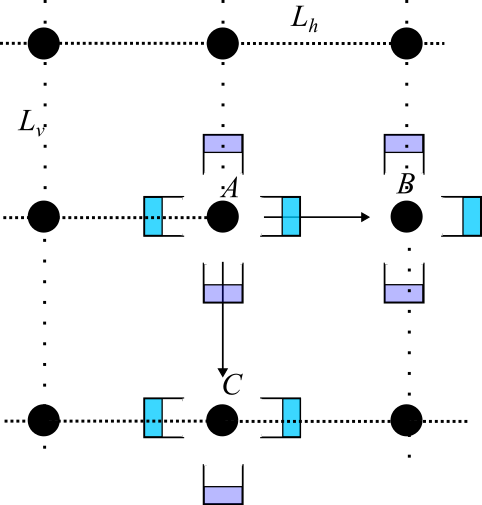}
	\caption{Simplified mesh of nodes}
	\label{fig:node_queues}
\end{figure}

The performance of the probabilistic routing algorithm can be analyzed in a simplified network as shown in Fig. \ref{fig:node_queues}. We use the simplifying assumptions that the network is an infinite mesh with the property that $L_h<L_v$, and that each node is fed packets with the same rate. Each node is associated with four queues, corresponding to the four ISLs a satellite can have.
For packets arriving at a node, the next hop could be in any direction. According to DRA, a primary and secondary direction is given to the congestion control algorithm, which then decides which direction to send the packet in.

\begin{figure}
	\centering
	\includegraphics[width=.7\columnwidth]{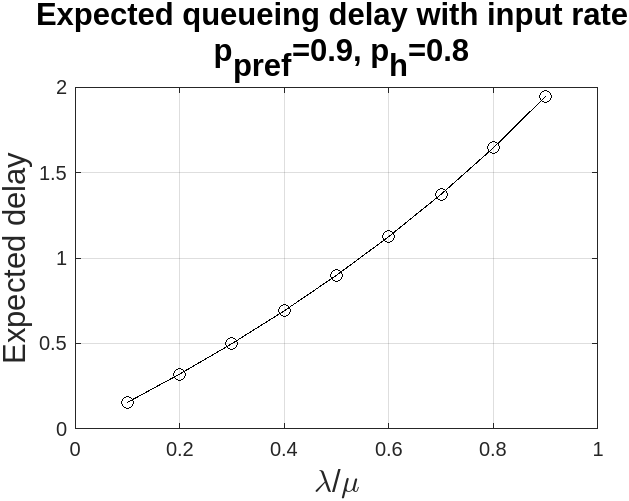}
	\caption{Variation of normalized expected queueing delay for a path with 3 horizontal hops and 3 vertical hops with $\lambda/\mu$}
	\label{fig:exp_delay_analytic}
\end{figure}

As seen in the routing algorithm, if the horizontal ISLs are shorter than the vertical ISLs, even if the source-destination pair distribution is uniform, the horizontal hops can be chosen as primary more often than the vertical hops, owing to the fact that the DRA tries to have its horizontal hops in the horizontal ring where the length of the horizontal ISLs is lower than in the rings that the path might encounter in its vertical hops. We assume that $p_h$ is the probability of choosing the primary direction to be horizontal, i.e, \textit{right} or \textit{left}. The choice of the secondary direction then comes down to two options, orthogonal to the primary direction, as the packet does not reverse its direction. Thus, for a packet, we can assume that
\begin{align*}
	P(\text{right primary, up secondary}) &= \frac{p_h}{2} \times \frac{1}{2} \\
	P(\text{up primary, right secondary}) &= \frac{1-p_h}{2} \times \frac{1}{2}
\end{align*} 
According to the probabilistic congestion control algorithm, node A has the latest traffic metrics from nodes B and C stored. Assuming that the traffic metric sent by $node$ is purely the average of the outgoing queue lengths, i.e, $w_{neighbor}=0$ in \eqref{eq:weighted_sum}, the traffic metrics calculated by B and C to be sent to A are 
\begin{align}
	m_{B\to A} &= \frac{N_{B, right} + N_{B, up} + N_{B, down}}{3}\\
	m_{C\to A} &= \frac{N_{C, right} + N_{C, down} + N_{C, left}}{3}
\end{align}
where $N_{node, direction}$ is the instantaneous length of the queue in node $node$ towards $direction$. Assuming that the mesh is infinite, and a uniform distribution of source-destination flows, the mean queue length is the same along the horizontal directions, and similarly along the vertical directions, and across nodes. 
Thus,
\begin{align}
	E(N_{B, right}) = E(N_{B, left}) &= E(N_{C,right})= N_h\\
	E(N_{B, up}) = E(N_{B, down}) &= N_v 
\end{align} 
Thus, the expected values of the traffic metrics are 
\begin{align*}
	E(m_{B\to A}) = m_{h} &= \frac{N_h + 2N_v}{3}\\
	E(m_{C \to A}) = m_v &= \frac{N_v +2N_h}{3}  
\end{align*}
Assuming that $w_{buffer} = 0 $ in \eqref{eq:w_buffer}, the congestion levels compared are purely the traffic metrics sent by the nodes. If DRA gives \textit{right} as the primary direction for the packet at A, and \textit{down} as the secondary, the probability of going right is 
\begin{equation}
P(right |\text{right primary, down secondary}) = f_p(m_h, m_v, p_{pref}).
\end{equation}
Thus, for a packet, the probability of going right is 
\begin{align*}
	P(right) &= P(\text{right primary, down secondary})f_p(m_h, m_v, p_{pref})\\
	&+ P(\text{right primary, up secondary})f_p(m_h, m_v, p_{pref}) \\
	&+ P(\text{up primary, right secondary})f_p(m_v, m_h, p_{pref}) \\
	&+ P(\text{down primary, right secondary})f_p(m_v, m_h, p_{pref}) 
\end{align*}
After simplifying, the expression reduces to 
\begin{multline}
\label{eq:p_right_exp}
P(right) = \frac{p_{pref}p_h(N_v + 2N_h)}{N_h(1+p_{pref})+N_v(2-p_{pref})}\frac{1}{2} \\
	+ \frac{(1-p_h)(1-p_{pref})(N_v + 2N_h)}{N_v(1+p_{pref}) + N_h(2-p_{pref})}\frac{1}{2}.
\end{multline}
Now, assuming that the input rate to node A is $\lambda$, the input rate to the queue to the right of node A is $\lambda P(right)$. Assuming that all queues are M/M/1\cite{mm1_stochastic} with service rate as $\mu$, and that all queues are stable,
the mean queue length of the queue to the right of node A is 
\begin{equation}
\label{eq:p_right}
	N_h = \frac{\rho_h}{1-\rho_h}
\end{equation}
where $\rho_h = \lambda P(right)/\mu$ (the number of packets in the queue follows a geometric distribution with parameter $1-\rho$ \cite{harchol-balter_2013}). $P(up)$ is calculated as in \eqref{eq:p_right_exp}, and gives 
\begin{equation}
\label{eq:p_up}
	N_v = \frac{\rho_v}{1-\rho_v}
\end{equation}
where $\rho_v = \lambda P(up)/\mu$. Thus, \eqref{eq:p_right} and \eqref{eq:p_up} form a system with two variables, namely $N_v$ and $N_h$. This system of multivariate polynomial equations was solved using MATLAB. For a path involving 3 horizontal hops and 3 vertical hops, the expected queuing delay can be expressed as $(3N_h + 3N_v)t_{tx}$, where $t_{tx}$ is the transmission delay of a single packet, assuming that the optimal path is followed. The variation of this queuing delay (normalized with $t_{tx}$) with $\lambda$ can be seen in Fig. \ref{fig:exp_delay_analytic}.
\section{Simulation Setup and Results}
\label{sec:setup}
Simulations were done in Python 3. A constellation with 12 polar orbital planes and 24 satellites per plane was used. Thus, $N=12$ and $M=24$ were the parameters for the constellation, with an inclination of $90^\degree$ and an altitude of 600 km. A discrete event simulator was built which executed events in order of lowest time of execution. In the plot, each data point was obtained by averaging over 20 runs.

A portion of the network was simulated, bounded by the nodes $(2,3), (7,3), (7,9), (2,9)$, consisting of 42 nodes. The simulation was fed with Poisson arrivals with rate $\lambda_{in}$ in each flow. After every $t_{step}$ seconds, $n_{pairs}$ source-destination pairs were chosen uniformly randomly from the 42 nodes, and $n_{packets}$ packets were queued up for these pairs with exponential inter-arrival times. Then transmissions were stopped, and the network was allowed to decongest. 1 kB packets were used with a transmission rate of 25 Mbps. The buffer size was chosen to be 200, and the threshold to be 150. Parameters chosen via simulations are $p_{preference}=0.9$, $w_{ngbr}=0.25$, $w_{buffer}=0.8$ to minimize average end-to-end delay.


\begin{figure}
	\centering
	\begin{subfigure}[b]{0.7\columnwidth}
		\includegraphics[width=\columnwidth]{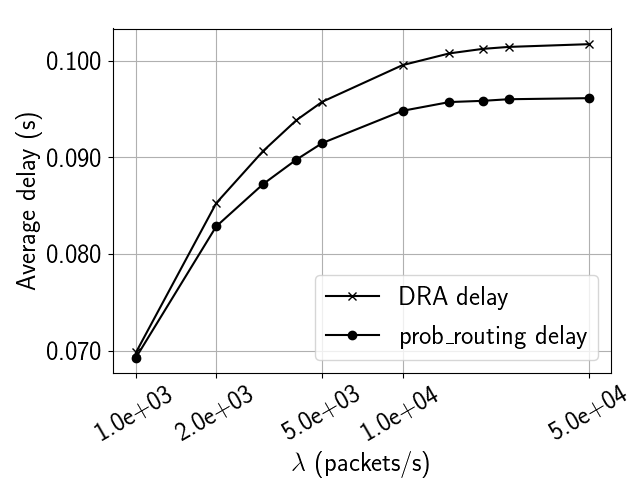}
		\caption{Variation with $\lambda_{in}$, for all flows}
		\label{fig:all_flows_delay_lamda_var}
	\end{subfigure}
	\begin{subfigure}[b]{0.7\columnwidth}
	\includegraphics[width=\columnwidth]{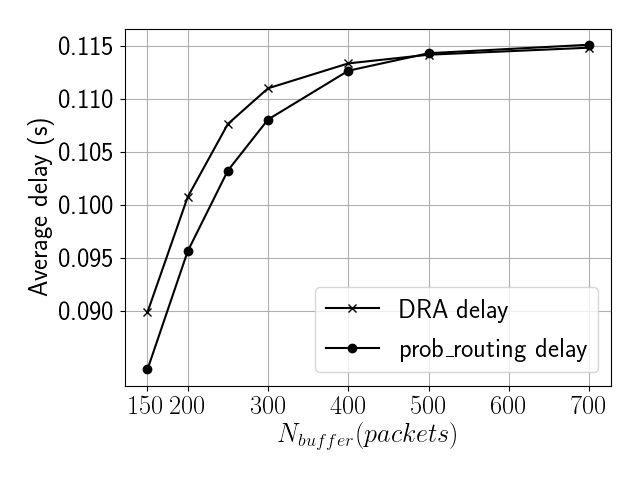}
	\caption{Variation with $N_{buffer}$, for all flows}
	\label{fig:all_flows_delay_drop_buff_var}		
	\end{subfigure}
\caption{Comparison of DRA and proposed congestion control algorithms}
\end{figure}
The proposed probabilistic routing algorithm performs better than the DRA in terms of average end-to-end delay for higher input rates ($\lambda_{in}$) to the network (Fig. \ref{fig:all_flows_delay_lamda_var}). For $\lambda_{in} = 1.5\times 10^4$ packets/s, the probabilistic routing algorithm gives an improvement of 5.041 ms, which is significant as the propagation delay in one hop is of the order of 6 ms.
When $N_{buffer}$ is changed, the buffer sizes in each node are increased. The ratio $N_{threshold}/N_{buffer}$ is fixed to be $0.75$, to study the effect of changing $N_{buffer}$ primarily. For large buffers, both algorithms perform similarly as no node gets congested. For lower buffer sizes, the proposed scheme performs better than the DRA in terms of end-to-end delay. Fig. \ref{fig:all_flows_delay_drop_buff_var} shows the variation of average end-to-end delay with $N_{buffer}$. For $N_{buffer} = 200$, the probabilistic routing algorithm gives an improvement of 5.118 ms over the DRA.

\section{Conclusions and Future Work}
\label{sec:conclusions}
The proposed probabilistic congestion control algorithm offers lower average end-to-end delay in congested networks, as compared to the existing DRA congestion control algorithm. It uses the regular nature of the topology to exchange information, and effectively route packets in a probabilistic manner. For now, this scheme does not account for satellite or link failures. This could be done by flooding the network, or by incorporating a different header into the packets being sent by neighboring nodes to indicate local link failures. Loop free routing can be implemented by maintaining a partial list of the nodes visited by the packet (in $(p,s)$ format) without significant overhead. 
\pagebreak
\bibliography{refs.bib}
\bibliographystyle{ieeetr}
\end{document}